\documentclass[aps,twocolumn,showpacs]{revtex4}
\usepackage{amsmath,amssymb}
\usepackage{graphics,graphicx}
\usepackage{dcolumn,bm}

\begin{document}
\title{Ideal-gas like market models with savings: quenched and annealed cases}

\author{
Arnab Chatterjee
}
\email{arnab.chatterjee@saha.ac.in}
\author{
Bikas K. Chakrabarti
}
\email{bikask.chakrabarti@saha.ac.in}
\affiliation{Theoretical Condensed Matter Physics Division and 
Centre for Applied Mathematics and Computational Science, Saha Institute of Nuclear Physics, 1/AF Bidhannagar, Kolkata 700064, India.}

\begin{abstract}
We analyze the ideal gas like models of markets and
review the different cases where a `savings' factor changes
the nature and shape of the distribution of wealth. 
These models can produce similar distribution of wealth
as observed across varied economies.
We present a more realistic model where the saving factor can vary over 
time (annealed savings) and yet produces Pareto distribution
of wealth in certain cases. We discuss the relevance of such models
in the context of wealth distribution, and address some recent
issues in the context of these models.
\end{abstract}

\pacs{89.20.Hh,89.75.Hc,89.75.Da,43.38.Si}
\maketitle
\section{Introduction}
The study of wealth distribution~\cite{cc:EWD05} in a society has 
remained an intriguing problem since Vilfredo Pareto who first
observed~\cite{cc:Pareto:1897} that the number of rich people with 
wealth $m$ decay following an inverse:
\begin{equation}
P(m) \sim m^{-(1+\nu)}.
\label{par}
\end{equation}
$P(m)$ is number density of people possessing wealth $m$,
and $\nu$ is known as the Pareto exponent. This exponent generally
assumes a value between $1$ and $3$ in varied 
economies~\cite{cc:realdatag,cc:realdataln,cc:Sinha:2006}.
It is also known that for low and medium income, the number density
$P(m)$ falls off much faster: exponentially~\cite{cc:realdatag} or
in a log-normal way~\cite{cc:realdataln}.

In recent years, easy availability of data has helped in the
analysis of wealth or income distributions in various societies~\cite{cc:EWD05}.
It is now more or less established that the distribution
has a power-law tail for the large (about 5\% of the population)
wealth/income~\cite{cc:realdatag,cc:realdataln,cc:Sinha:2006},
while the majority (around 95\%) low income distribution fits well to
Gibbs or log-normal form.

There has been several attempts to model a simple economy
with minimum trading ingredients, 
which involve a wealth exchange process~\cite{cc:othermodels}
that produce a distribution
of wealth similar to that observed in the real market.
We  are particularly interested in microscopic models of markets where
the (economic) trading activity is considered as a scattering
process~\cite{cc:marjit,cc:Dragulescu:2000,cc:Chakraborti:2000,cc:Hayes:2002,cc:Chatterjee:2004,cc:Chatterjee:2003,cc:Chakrabarti:2004,cc:Slanina:2004}
(see also Ref.~\cite{cc:ESTP:KG} for a recent extensive review).
We concentrate on models that incorporate `savings' as an essential 
ingredient in a trading process, and reproduces the salient features 
seen across wealth distributions in varied economies
(see Ref.~\cite{cc:EWD:CC} for a review).
Angle~\cite{cc:Angle:1986} studied inequality processes which
can be mapped to the savings wealth models is certain 
cases; see Ref.~\cite{cc:Angle:2006} for a detailed review.

These studies also show (and discussed briefly here) 
how the distribution of savings can
be modified to reproduce the salient features of empirical
distributions of wealth -- namely the shape of the distribution
for the low and middle wealth and the tunable Pareto exponent.
In all these 
models~\cite{cc:Chakraborti:2000,cc:Hayes:2002,cc:Chatterjee:2004,cc:Chatterjee:2003,cc:Chakrabarti:2004},
`savings' was introduced as an annealed parameter that
remained invariant with time (or trading). 

Apart from a brief summary of the established results in such models,
here we report some new results for annealed cases, 
where the saving factor can change with time, as one would expect
in a real trading process. We report cases where the wealth distribution
is still described by a Pareto law. We also forward some justification
of the various assumptions in such models.

\section{Ideal-gas like models of trading}
We first consider an ideal-gas model of a closed economic system.
Wealth is measured in terms of the amount of money possessed by an
individual.
Production is not allowed i.e, total money $M$ is fixed and  also there
is no migration of population i.e, total number of agents $N$ is fixed, 
and the only economic activity is confined to trading.
Each agent $i$, individual or corporate, possess money $m_i(t)$ at time $t$.
In any trading, a pair of agents $i$ and $j$ randomly exchange their 
money~\cite{cc:marjit,cc:Dragulescu:2000,cc:Chakraborti:2000}, 
such that their total money is (locally) conserved
and none end up with negative money ($m_i(t) \ge 0$, i.e, debt not allowed):
\begin{equation} 
\label{consv}
m_i(t) + m_j(t) = m_i(t+1) + m_j(t+1);
\end{equation}
time ($t$) changes by one unit after each trading.
The steady-state ($t \to \infty$) distribution of money is Gibbs one:
\begin{equation}
\label{gibbs}
P(m)=(1/T)\exp(-m/T);T=M/N. 
\end{equation} 
No matter how uniform or justified the initial distribution is, the
eventual steady state corresponds to Gibbs distribution where most of the
people end up with very little money.
This follows from the conservation of money and additivity of entropy:
\begin{equation}
\label{prob}
P(m_1)P(m_2)=P(m_1+m_2).
\end{equation}
This steady state result is quite robust and realistic.
Several variations of the trading, and of the `lattice'
(on which the agents can be put and each agent trade with its
`lattice neighbors' only) --- compact, fractal or small-world 
like~\cite{cc:EWD05}, does not affect the distribution. 

\section{Savings in Ideal-gas trading market: Quenched case}
In any trading, savings come naturally~\cite{cc:Samuelson:1980}.
A saving factor $\lambda$ is therefore introduced in the same 
model~\cite{cc:Chakraborti:2000} (Ref.~\cite{cc:Dragulescu:2000} 
is the model without savings), where each trader
at time $t$ saves a fraction $\lambda$ of its money $m_i(t)$ and trades
randomly with the rest.
In each of the following two cases, the savings fraction does
not vary with time, and hence we call it `quenched' in the terminology
of statistical mechanics.

\subsection{Fixed or uniform savings}
For the case of `fixed' savings, the money exchange rules are:
\begin{equation}
\label{delm}
m_{i}(t+1)=m_{i}(t)+\Delta m; \  m_{j}(t+1)=m_{j}(t)-\Delta m 
\end{equation}
\noindent where
\begin{equation}
\label{eps}
\Delta m=(1-\lambda )[\epsilon \{m_{i}(t)+m_{j}(t)\}-m_{i}(t)],
\end{equation}
where $\epsilon$ is a random fraction, coming from the stochastic nature
of the trading. $\lambda$ is a fraction ($0 \le \lambda < 1$)
which we call the saving factor.

The market (non-interacting at $\lambda =0$ and $1$) becomes `interacting'
for any non-vanishing $\lambda (<1)$: For fixed $\lambda$ (same for all
agents), the steady state distribution $P_f(m)$ of money is sharply
decaying on both sides with the most-probable money per agent shifting away
from $m=0$ (for $\lambda =0$) to $M/N$ as $\lambda \to 1$.
The self-organizing feature of this market,
induced by sheer self-interest of saving by each agent without any global
perspective, is very significant as the fraction of paupers decrease with
saving fraction $\lambda$ and most people possess some fraction of the
average money in the market (for $\lambda \to 1$, the socialists'
dream is achieved with just people's self-interest of saving!).
Although this fixed saving propensity does not give the Pareto-like
power-law distribution yet, the Markovian nature of the scattering or trading
processes (eqn.~(\ref{prob})) is lost and the system becomes co-operative.
Indirectly through $\lambda$, the agents get to develop a correlation with
(start interacting with) each other and the system co-operatively 
self-organizes~\cite{cc:Bak:1997} towards a most-probable distribution.

This model has been understood to a certain extent 
(see e.g,~\cite{cc:Das:2003,cc:Patriarca:2004,cc:Repetowicz:2005}), 
and argued to resemble a gamma distribution~\cite{cc:Patriarca:2004}, 
and partly explained analytically.
This model clearly finds its relevance in cases where the economy consists
of traders with `waged' income~\cite{cc:Willis:2004}.

\subsection{Distributed savings}
In a real society or economy, $\lambda$ is a very inhomogeneous parameter:
the interest of saving varies from person to person.
We move a step closer to the real situation where saving factor $\lambda$ is
widely distributed within the 
population~\cite{cc:Chatterjee:2004,cc:Chatterjee:2003,cc:Chakrabarti:2004}.
The evolution of money in such a trading
can be written as:
\begin{equation}
\label{mi}
m_i(t+1)=\lambda_i m_i(t) + \epsilon_{ij} \left[(1-\lambda_i)m_i(t) + (1-\lambda_j)m_j(t)\right], 
\end{equation}
\begin{equation}
\label{mj}
m_j(t+1)=\lambda_j m_j(t) + (1-\epsilon_{ij}) \left[(1-\lambda_i)m_i(t) + (1-\lambda_j)m_j(t)\right]
\end{equation}
One again follows the same rules as before, except that
\begin{equation}
\label{lrand}
\Delta m=\epsilon_{ij}(1-\lambda_{j})m_{j}(t)-(1-\lambda _{i})(1 - \epsilon_{ij})m_{i}(t)
\end{equation}
here; $\lambda _{i}$ and $\lambda _{j}$ being the saving
propensities of agents $i$ and $j$. The agents have fixed (over time) saving
propensities, distributed independently, randomly and uniformly (white)
within an interval $0$ to $1$ agent $i$ saves a random fraction
$\lambda_i$ ($0 \le \lambda_i < 1$) and this $\lambda_i$ value is quenched
for each agent ($\lambda_i$ are independent of trading or $t$).
$P(m)$ is found to follow a strict power-law decay.
This decay fits to Pareto
law (\ref{par}) with $\nu = 1.01 \pm 0.02$ for several decades.
This power law is extremely robust: a distribution 
\begin{equation}
\label{lam0}
\rho(\lambda) \sim |\lambda_0-\lambda|^\alpha, 
\ \ \lambda_0 \ne 1, \ \ 0 \le \lambda<1,
\end{equation}
of quenched $\lambda$ values among the agents produce power law
distributed $m$ with Pareto index $\nu=1$, irrespective of the value of
$\alpha$. For negative $\alpha$ values, however,
we get an initial (small $m$) Gibbs-like decay in $P(m)$.
In case $\lambda_0 =1$, the Pareto exponent is modified to
$\nu=1+\alpha$, which qualifies for the non-universal exponents
in the same model \cite{cc:Chatterjee:2004,cc:Mohanty:2006}.

This model~\cite{cc:Chatterjee:2004} has been thoroughly analyzed, 
and the analytical
derivation of the Pareto exponent has been achieved in certain 
cases~\cite{cc:Repetowicz:2005,cc:Mohanty:2006,cc:Chatterjee:2005}.
The Pareto exponent has been derived to exactly $1$.

In this model, agents with higher saving propensity tend to hold 
higher average wealth, which is justified by the fact that the saving
propensity in the rich population is always high~\cite{cc:Dynan:2004}.

\section{Savings in Ideal-gas trading market: Annealed case}

\begin{figure}[t]
\includegraphics[width=8.5cm]{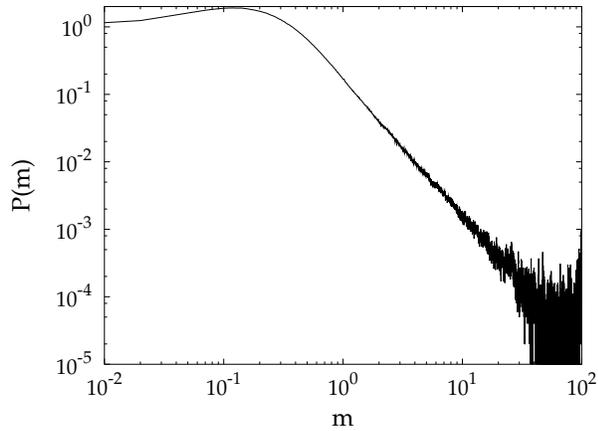}
\caption{
Distribution $P(m)$ of money $m$ in case of annealed savings $\lambda$
varying randomly in $[\mu,1)$. Here, $\zeta(\mu)$ has a uniform distribution.
The distribution has a power law tail with Pareto index $\nu=1$.
The simulation has been done for a system of $N=10^2$ agents, with $M/N=1$.
$P(m)$ is the steady state distribution after $4 \times 10^4 \times N$
random exchanges, and averaged over an ensemble of $10^5$. 
}
\label{fig:ann:1-lambda}
\end{figure}
In a real trading process, the concept of `saving factor' cannot be attributed
to a quantity that is invariant with time. A saving factor always
changes with time or trading.
In earlier works, we reported the case of annealed savings, 
where the savings factor $\lambda_i$ changes with time in the
interval $[0,1)$, but does not produce a power law in 
$P(m)$~\cite{cc:Chatterjee:2004}. We report below some special cases 
of annealed saving which produce a power law distribution of $P(m)$.

If one allows the saving factor $\lambda_i$ to vary with time
in $[0,1)$, the money distribution $P(m)$ does not produce a
power law tail.

Instead, we propose a slightly different model of an annealed saving case.
We associate a parameter $\mu_i$ ($0 < \mu_i < 1$) with each agent $i$
such that the savings factor $\lambda_i$ randomly assumes a value in the
interval $[\mu_i,1)$ at each time or trading.
The trading rules are of course unaltered and governed by Eqns.~(\ref{mi})
and (\ref{mj}).
Now, considering a suitable distribution $\zeta(\mu)$ of $\mu$
over the agents, one can produce money distributions with power-law
tail. The only condition that needs to hold is that  $\zeta(\mu)$
should be non-vanishing as $\mu \to 1$.
Figure~\ref{fig:ann:1-lambda} shows the case when $\zeta(\mu)=1$.
Numerical simulations suggest that the behavior of the wealth 
distribution is similar to the quenched savings case. In other
words, only if $\zeta(\mu) \propto |1-\mu|^\alpha$, it is reflected
in the Pareto exponent as $\nu=1+\alpha$.

\section{Relevance of gas like models}
Al these gas-like models of trading markets are based on the assumption of
(a) money conservation (globally in the market; as well as locally 
in any trading) 
and 
(b) stochasticity.
These points have been criticized strongly (by economists) in the 
recent literature~\cite{cc:Gallegati:2006}.
In the following, we forward some of the arguments in favour of these
assumptions (see also~\cite{cc:ESOM:2006}).

\subsection{Money conservation}
If we view the trading as scattering processes, one can see the 
equivalence. Of course, in any such `money-exchange' trading process,
one receives some profit or service from the other and this does
not appear to be completely random, as assumed in the models.
However, if we concentrate only on the `cash' exchanged 
(even using Bank cards!), every trading is a money conserving one 
(like the elastic scattering process in physics!)

It is also important to note that the frequency of money exchange
in such models define a time scale in which the total money
in the market does not change. In real economies, the total money
changes much slowly, so that in the time scale of exchanges,
it is quite reasonable to assume the total money to be conserved in
these exchange models. This can also be justified by the fact that
the average international foreign currency exchange rates change
drastically (say, by more than 10\%) very rarely; according to the
Reserve Bank of India, the US Dollar remained at INR $45 \pm 3$ for
the last eight years~\cite{cc:Sarkar:2006}! 
The typical time scale of the exchanges 
considered here correspond to seconds or minutes and hence the constancy
assumption cannot be a major problem.

\subsection{Stochasticity}
But, are these trading random? Surely not, when looked upon from
individual's point of view: When one maximizes his/her utility by
money exchange for the $p$-th commodity, he/she may choose to
go to the $q$-th agent and for the $r$-th commodity he/she will go to the
$s$-th agent. But since $p \ne q \ne r \ne s$ in general, when viewed
from a global level, these trading/scattering events will all look
random (although for individuals these is a defined choice or utility
maximization).

Apart from the choice of the agents for any trade, the traded amount
are considered to be random in such models. Some critics argue, this
cannot be totally random as the amount is determined by the price of the
commodity exchanged. Again, this can be defended very easily. If a little
fluctuation over the `just price' occurs in each trade due to the bargain
capacities of the agents involved, one can easily demonstrate that after
sufficient trading (time, depending on the amount of fluctuation 
in each trade), the distribution will be determined by the stochasticity, 
as in the cases of directed random walks or other biased stochastic models
in physics.

It may be noted in this context that in the stochastically formulated
ideal gas models in physics (developed in late 1800/early 1900) 
one (physicists) already knew for more than a hundred years, that
each of the constituent particle (molecule) follows a precise equation
of motion, namely that due to Newton.
The assumption of stochasticity here in such models, even though each agent
might follow an utility maximizing strategy (like Newton's equation
of motion for molecules), is therefore, not very unusual in the context.

\section{Summary and conclusions}
We analyze the gas like models of markets.
We review the different cases where a quenched `savings' factor changes
the nature and shape of the distribution of wealth. Suitable modification
in the nature of the `savings' distribution can simulate all observed
wealth distributions. 
We give here some new numerical results for the annealed `savings' case.
We find that the more realistic model, where the saving factor 
randomly varies in time (annealed savings), still produce a Pareto distribution
of wealth in certain cases.
We also forward some arguments in favour of the assumptions made in
such gas-like models.




\end{document}